\documentclass{emulateapj} 
\usepackage{apjfonts}
\usepackage{epsfig,graphicx}

\newcommand{\sci}{{\small I}\ }

\slugcomment{Received ........; Accepted .........}

\shorttitle{Can overturning motions in penumbral filaments be detected?}

\shortauthors{Bharti et al.}

\begin{document}

\title{Can overturning motions in penumbral filaments be detected?}

\author{Lokesh Bharti$^1$}
\author{Manfred Sch{\"u}ssler$^1$}
\author{Matthias Rempel$^2$}
\affil{$^1$Max-Planck-Institut f\"ur Sonnensystemforschung,
       Max-Planck-Str. 2,
       37191 Katlenburg-Lindau, Germany,}
\affil{$^2$High Altitude Observatory, NCAR, P.O. Box 3000,
       Boulder, CO 80307, USA}
       \email{bharti@mps.mpg.de}

\begin{abstract}
Numerical simulations indicate that the filamentation of sunspot
penumbrae and the associated systematic outflow (the Evershed effect)
are due to convectively driven fluid motions constrained by the
inclined magnetic field. We investigate whether these motions, in
particular the upflows in the bright filaments and the downflows at
their edges can be reliably observed with existing instrumentation. We
use a snapshot from a sunspot simulation to calculate 2D maps of
synthetic line profiles for the spectral lines Fe\sci 7090.4 \AA~ and
C\sci 5380.34 \AA. The maps are spatially and spectrally degraded
according to typical instrument properties. Line-of-sight velocities
are determined from line bisector shifts.  We find that the
detectability of the convective flows is strongly affected by spatial
smearing, particularly so for the downflows.  Furthermore, the
line-of-sight velocities are dominated by the Evershed flow unless the
observation is made very near to disk center.  These problems may
  have compromised recent attempts to detect overturning penumbral
convection. Lines with a low formation height are best suited to
detect the convective flows.
\end{abstract}

\keywords{Sun: convection -- sunspots -- Sun:
  magnetic field}

\section{Introduction}
\label{sec:intro}

Significant progress in understanding the physical processes
underlying the photospheric structure and dynamics of sunspots has
been made in recent years by the interplay of observational results
obtained with advanced instrumentation on the ground and in space with
the results of comprehensive numerical simulations. In particular,
simulations indicate that magneto-convective processes dominate the
dynamics and structure formation in the umbra (Sch\"ussler \& V\"ogler
2006, Bharti et al. 2010a) and penumbra (Heinemann et al. 2007,
Scharmer et al. 2008, Rempel et al. 2009ab, Rempel 2011). Sunspot
fine structure seen in the photosphere, such as umbral dots, light
bridges, and penumbral filaments, are attributed to overturning
convection in the presence of a vertical or inclined magnetic field.

Observational evidence has been found in support of the notion of
magneto-convection in the umbra (Bharti et al. 2007ab, 2009, Ortiz et
al. 2010) and in light bridges (Bharti et al. 2007c, Rimmele 2008,
Rouppe van der Voort et al. 2010).  Observational support for
convection in penumbral filaments is not so clearcut, although some
indirect evidence has been found by M\'arquez et al. (2006) and
S\'anchez Almeida et al. (2007). From numerical models there are two
aspects to this flow that are manifest in observations: Upflows in
bright filaments in the inner penumbra lead to dark cores and
particularly bright filament heads moving toward the umbra during
their formation phase (Rempel et al. 2009a); along the full length of
filaments upflows are deflected by the Lorentz force due to the
inclined magnetic field and turn into the almost horizontal Evershed
flows along the filaments near the surface of optical depth unity
(Rempel 2011). Part of the upflowing material, however, turns over in
a roll-type motion perpendicular to the direction of the filament axis
and descends in downflows along the edges of the filaments. Indirect
evidence for this kind of overturning flows is provided by the
observations of `twisting' motions (Ichimoto et al. 2007, Bharti et
al. 2010b) in penumbral filaments. The associated upflow velocities
seem to be high enough to balance the radiative losses of filaments
over an extended part of their length (Zakharov et al. 2008, Bharti et
al. 2010b).  On the other hand, the associated downflows along the
edges of penumbral filaments have not yet been clearly detected by
spectroscopic observations (e.g., Franz \& Schlichenmaier 2009,
Katsukawa \& Jurcak 2010, Louis et al. 2011).  For instance, Bellot Rubio et al. (2010)
used high-resolution observations of the line Fe I 7090.4 \AA~ line
with the Swedish 1-m telescope (Scharmer et al. 2003) on La Palma and
found no such downflows above their detection limit of a few hundred
m$\,$s$^{-1}$. In fact, the vertical penumbral velocity field derived
by these authors appears to be dominated by upflows along the line of
sight.

The aim of this paper is to study, using the most advanced numerical
simulations of sunspots, under which conditions overturning convective
flows in penumbral filaments of the kind present in the simulations
would be detectable in observations. Thereby, we wish to clarify whether
negative results such as those of Bellot Rubio et al. (2010) are in fact
in contradiction to those of the numerical simulations.

\section{Simulations, line synthesis and convolution}
\label{sec:simul}
We consider a single snapshot from a sunspot simulation using the
MURaM code (for details, see V{\"o}gler et al. 2005, Rempel et
al. 2009b) with a grid resolution of 16 km in vertical and 32 km in
the horizontal directions. The dimensions of the computational box are
$49.152\times 49.152$~Mm$^2$ horizontally and 6.144~Mm in depth. While
this simulation shares the basic ingredients with Rempel et al. 2009b,
the numerical setup consists of an individual sunspot (instead of the
sunspot pair). To obtain extended penumbrae in presence of periodic
boundaries (imposing same polarity spots nearby), we artificially
increased the inclination angle at the top boundary (700~km above the
quiet sun photosphere) by increasing the horizontal field components
by a factor of two compared to a potential field. The total (solar)
time elapsed since initialization of the simulation is about 6h. The
sunspot has a total flux of about $10^{22}$~Mx, the fraction of the
spot area covered by penumbra (defined through $0.5<I/I_{\odot}<0.9$)
is $65\%$. The first 3.3 hours were run in a resolution of 24~km in
vertical and 48~km in the horizontal directions, the following 2.7
hours were run in the high resolution of $16\times 32$~km$^2$. The
last 26 min were performed in addition with non-grey radiative
transfer using 4 opacity bins optimized for the quiet-Sun
conditions. This sunspot is part of a convergence study in which the
influence of the grid resolution on penumbral fine structure is
investigated (Rempel 2010, Rempel in prep.).

\begin{figure*}
\centering
\includegraphics[width=1\textwidth]{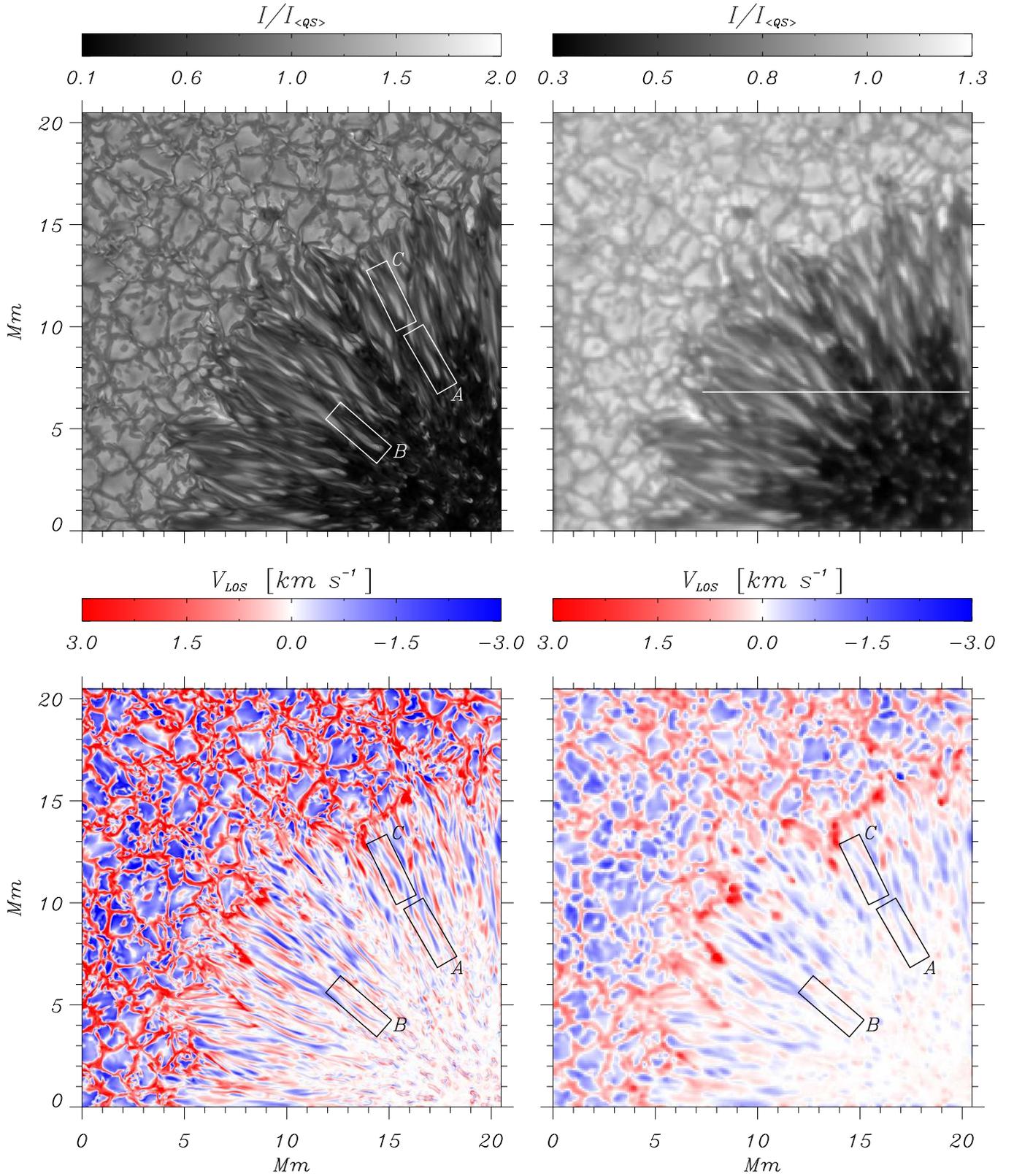}
\caption{Upper row: synthetic continuum intensity map for the Fe{\small
  I} 7090 \AA~ line at original simulation resolution (left) and at
  degraded resolution after convolution with a Gaussian profile of
  0$^{\prime\prime}$.25 FWHM (right). White rectangles indicate selected
  filaments that are studied in more detail in
  Figures~\ref{fig:figure5}-\ref{fig:figure6}. The horizontal line in the
  degraded image indicates the location of the artificial `slit' for the
  velocity profiles shown in Figure~\ref{fig:figure3}. Lower row:
  bisector velocity at 90 $\%$ relative intensity for the Fe{\small I}
  7090 \AA~ line at original simulation resolution (left) and at degraded
  resolution (right). The degraded images have been resampled to pixel
  size equal to half of the smearing FWHM.}
\label{fig:figure1}
\end{figure*}

For the analysis
carried out in this paper, we consider a $20.5\times20.5$~Mm$^2$
($640\times640$ pixel horizontally) section of the snapshot, covering
a quarter of the simulated sunspot (cf. Figure~\ref{fig:figure1}).
To determine synthetic line profiles we used the STOPRO (STOkes
PROfiles) code (Berdyugina et al. 2003), which calculates the Stokes
parameters for spectral lines in local thermodynamic equilibrium (LTE).
We obtained intensity (Stokes-$I$) profiles for the two lines
Fe\sci 7090.4 \AA~ and C\sci 5380.34 \AA~ for each of the $640\times640$
pixels of the simulation snapshot. The first of these lines has been
used in the study of Bellot Rubio et al. (2010), so that we can directly
compare with their results. The line is particularly well suited for
velocity measurements since its Land{\'e} factor is zero. The outer
wings of this line are formed around 100 km above optical depth unity
(Bellot Rubio et al. 2005). The C\sci 5380.34 \AA~ line forms in the deep
photosphere (Fleck 1991) and thus is more sensitive to the higher
convective velocities in these layers.

The maps of synthetic intensity profiles were considered a) at original
simulation resolution, and b) smeared and rebinned spatially and
spectrally according to observational and instrumental conditions.  The
point-spread function used for spatial convolution of the 7090 line maps
consists of a Gaussian core with $0^{\prime\prime}.25$ FWHM and
Lorentzian wings (cf. Shelyag et al. 2004) with an amplitude chosen such
that the RMS contrast of the continuum near to the line was degraded to
7\% outside the sunspot. The spectral smearing was carried out by
convolving with a Gaussian of 30~m\AA~ FWHM, corresponding to the
spectral resolution of the TRIPPEL spectrograph at the Swedish Solar
Telescope. The spatial and spectral resolutions of the degraded map of
intensity profiles for the 7090 line are therefore comparable to data
used by Bellot Rubio et al. (2010). For degrading the map of the
C\sci 5380.34 \AA~ line we used parameters corresponding to the CRISP 2D
spectropolarimeter at the SST: a Gaussian with $0^{\prime\prime}.14$
FWHM combined with a Lorentzian chosen to yield 10\% RMS continuum
contrast outside the sunspot for the spatial convolution and a Gaussian
with a FWHM of 40~m\AA~ for the convolution in wavelength.  All degraded
images have been resampled to (spatial and spectral) pixel sizes equal
to half of the FWHM of the corresponding Gaussian used for the
convolution.

Line-of-sight (LOS) Doppler velocities are computed by means of the line
bisector shifts for (relative) line intensity levels between 0$\%$ for
the line core and 100$\%$ for the continuum level, in steps of
10\%. Shifts for intermediate intensity levels are determined by linear
interpolation. Following standard observational procedure, we use the
darkest part of sunspot as zero velocity reference. According to the
varying formation height of the different parts of the line profile,
bisectors near the continuum sample the velocities at deeper layers than
those near the line core.

It should be kept in mind that the LOS speeds in our degraded
  maps still have to be considered as upper limits: we have assumed an
  ideal telescope and spectrograph, no noise and other instrumental
  effects. These can be quite important (cf. Joshi et al. 2011, see
  their Fig. 4 and Scharmer et al. 2011, see their Figs. S13-16).

Most of the results presented in the following section are for
synthetic line profiles calculated along vertical lines of sight,
corresponding to observations exactly at disk center. The effect of
a non-zero viewing angle, i.e., inclined lines of sight, is considered
in Sec.~\ref{subsec:incl}.

\begin{figure*}
\centering
\vspace{8mm}
\includegraphics[width=1\textwidth]{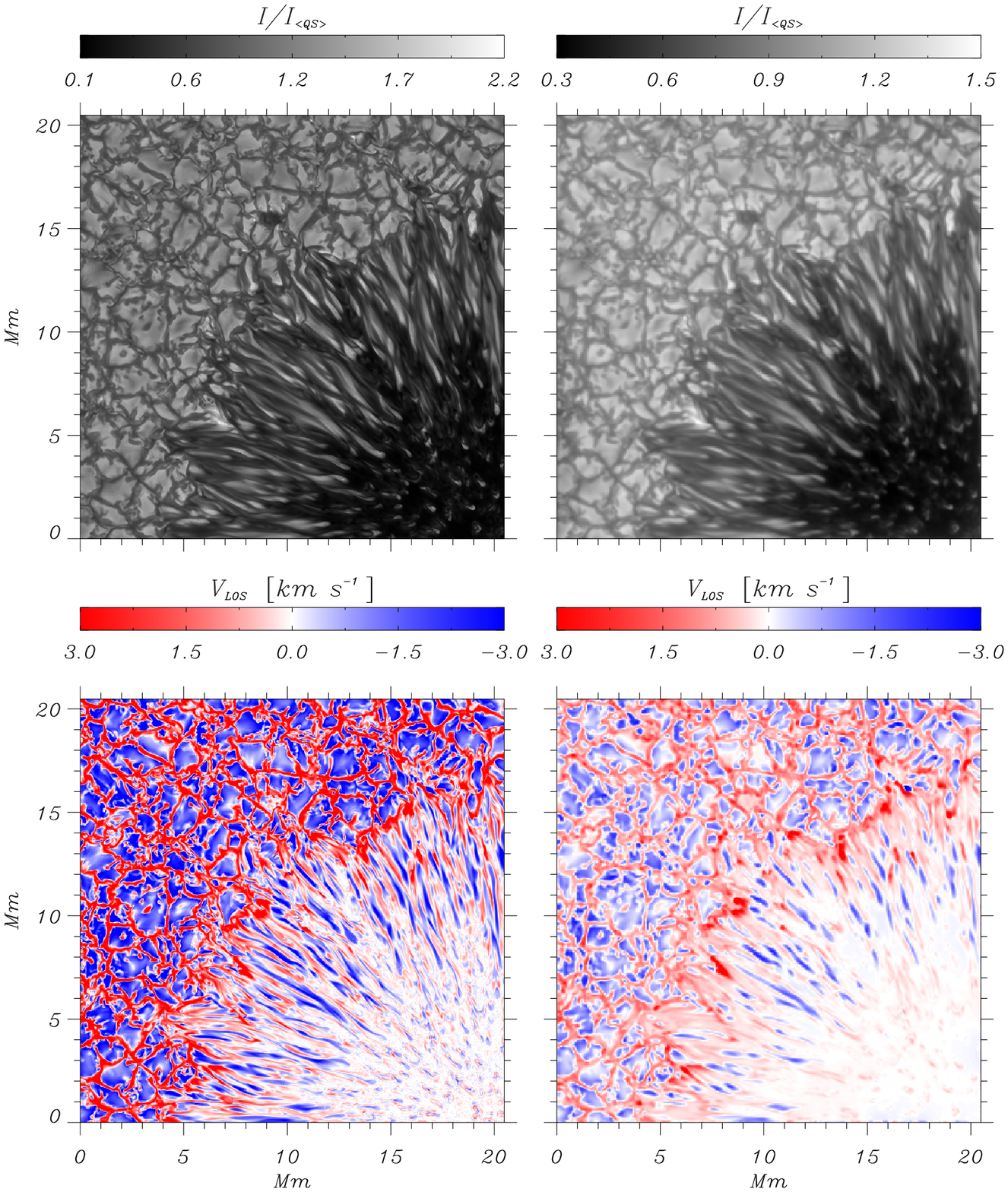}
\caption{Same as Figure~\ref{fig:figure1}, but for the line C{\small I}
  5380 \AA. The degraded images have been spatially convolved with a
  Gaussian of 0$^{\prime\prime}$.14 FWHM.}
\label{fig:figure2}
\end{figure*}

\section{Results}
\label{sec:results}
\subsection{Velocities from synthetic line profiles}
\label{subsec:velmaps}

The upper left panel of Figure~\ref{fig:figure1} shows a synthetic
continuum image for the 7090 \AA~ line at the original MHD simulation
grid resolution of 32~km. The granulation in the upper left part of the
image has an rms contrast of about 15\%.  White rectangles indicate the
penumbral filaments that are considered in more detail in
Sec.~\ref{subsec:filaments} below. The degraded and resampled continuum
image corresponding to a spatial resolution of 0$^{\prime\prime}.25$
resolution and 7\% rms contrast is shown in the upper right panel of
Figure~\ref{fig:figure1}. LOS velocity maps determined from the
bisectors at 90$\%$ intensity level are shown in the lower part of the
figure: the map on the left panel corresponds to the line profiles at
original resolution while the map on the right panel is based on the
(spatially and spectrally) degraded and resampled profiles. Note that
bisectors at 90$\%$ intensity levels can be strongly affected in
observational data by line blends and noise.  In our line synthesis none
of these effects is present.  Therefore, our results give an optimistic
estimate for the detectability of velocities based on line bisectors.
All vertical (LOS) velocities are significantly lower after degradation.
bf In particular, the downflow velocities at the edges of the
filaments are reduced from $\sim 1000$~m$\,$s$^{-1}$ to typically less
than $\sim 200$~m$\,$s$^{-1}$, so that they become much less visible in
the degraded map. Only at a few locations the downflow speed reaches
peak values of $\sim 300$~m$\,$s$^{-1}$. On the other hand, strong
downflow patches at the outer peripery of the penumbra remain clearly
detectable (e.g., Schlichenmaier \& Schmidt 1999, Franz \& Schlichenmaier
2009). These patches correspond to downturning Evershed flow in
filaments that re-enter the subsurface layers.

Figure~\ref{fig:figure2} shows the corresponding results for the 5380
\AA~ line using the 90$\%$ bisector. The continuum contrast is reduced
from 24\% at original
resolution (upper left panel) to 10\% in the degraded and resampled
image (upper right panel) corresponding to 0$^{\prime\prime}.14$
resolution. In comparison to the result for the 7090 \AA~ line, the
reduction of the LOS velocities in the degraded maps is less severe:
downflows at the edges of the penumbral filaments are reduced from
$\sim$2000~m$\,$s$^{-1}$ to m/s only to $\sim$800~m$\,$s$^{-1}$. Since the
flow velocities in the simulated penumbral filaments are highest around
optical depth unity (Rempel et al. 2009ab, Rempel 2011), these flows are better
sampled by the 5380 \AA~ line, owing to its lower formation height.

\begin{figure}
\hspace{5mm}
\includegraphics[width=0.44\textwidth,angle=0]{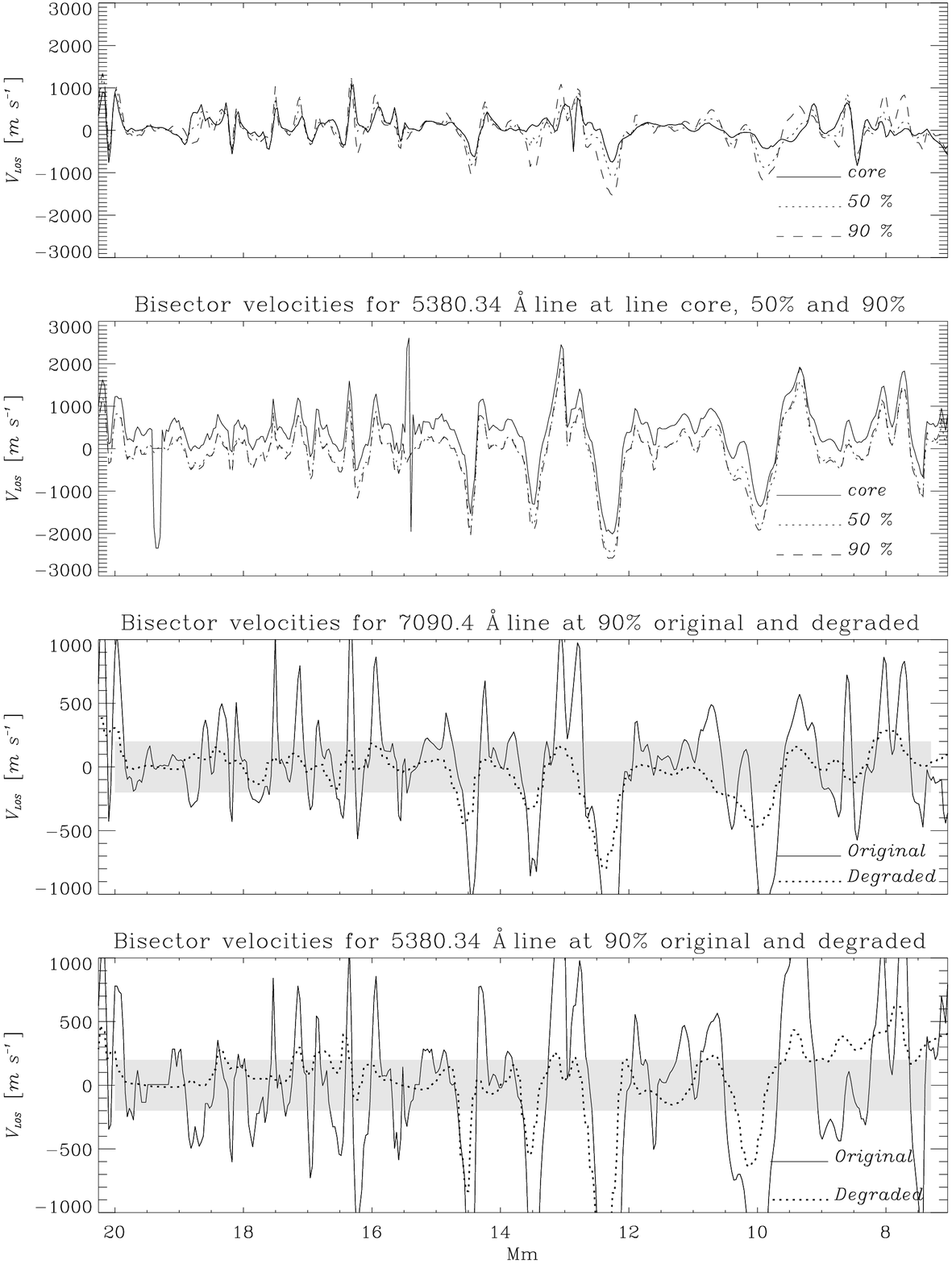}
\vspace{3mm}
\caption{Bisector velocities along the horizontal `slit' indicated in
  the upper left panel of Figure~\ref{fig:figure1}.  The gray stripes
  indicate the typical uncertainty of observational measurements
  $(\pm 200$~m$\,$s$^{-1}$).}
\label{fig:figure3}
\end{figure}

Figure~\ref{fig:figure3} shows LOS velocities along (from right to
left) the artificial `spectrograph slit' indicated by the white
horizontal line in the upper right panel of
Figure~\ref{fig:figure1}. The two upper panels illustrate, at original
simulation resolution, how well the depth-dependent velocities are
sampled by the two spectral lines.  As the vertical flow speeds
increase with depth in the atmosphere, the bisector shifts of the line
wings (at 50$\%$ and 90$\%$ relative intensity) show higher LOS speeds
than the line core shifts. Flow speeds determined from the deeper
originating 5380 \AA~ line are generally higher.  The lower two panels
of Figure~\ref{fig:figure3} illustrate the strong effect of spatial
and spectral smearing on the flow speeds obtained from the bisectors
at 90\% relative intensity. While the up- and downflows of the
overturning filament convection are conspicuous at original
resolution, the downflow speeds in the degraded maps rarely
exceed a typical observational detection limit of
$\pm200$~m$\,$s$^{-1}$ (Bellot Rubio et al. (2010) stated a detection
limit of $\sim50$~m$\,$s$^{-1}$ for their investigation with the
error in the position of the zero of the velocities about $\sim150$~m$\,$s$^{-1}$).
  On the other hand, the strongest upflows remain detectable.  Similar
  results are found for the slit positions at $y=10$~Mm and $y=12$~Mm
  (not shown here). Generally, the downflows are more affected by the
  image degradation than the upflows. This is probably due to the
  smaller spatial scales of the downflows and their preferred
  occurence in dark regions, so that they are more susceptible to
  spatial smearing and straylight. This has also been found by Joshi
  et al. (2011, see their Fig. 4) and Scharmer et al. (2011, see their Figs. S13-16).

\begin{figure}
\centering
\includegraphics[width=.5\textwidth]{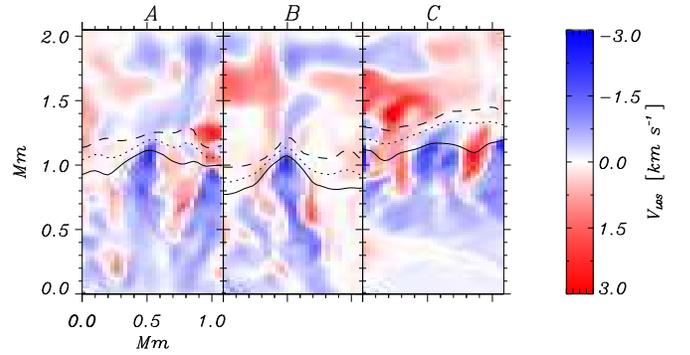}
\caption{Vertical velocities in the simulation. Shown are vertical
slices cut perpendicular to the direction of penumbral filaments A, B,
and C. The locations of the cuts are indicated on the intensity maps in
Figures~\ref{fig:figure5}, \ref{fig:figure7}, and \ref{fig:figure9},
respectively. The lines indicate levels of constant optical depth for
the continuum at 7090 \AA: $\tau=1$ (solid) $\tau=0.1$ (dotted), and
$\tau=0.01$ (dashed).}
\label{fig:figure4}
\end{figure}

\begin{figure}
\centering
\includegraphics[width=.5\textwidth,angle=0]{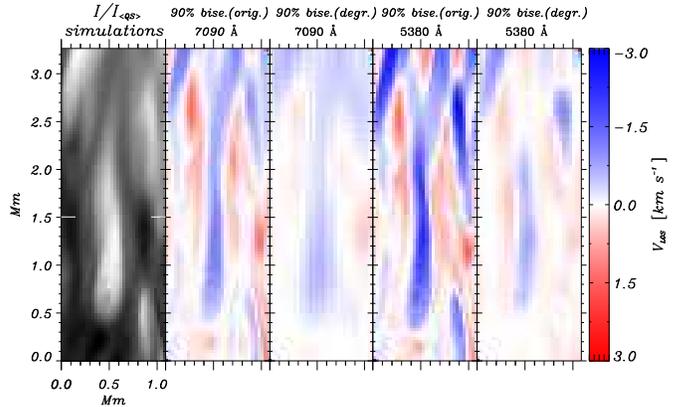}
\caption{Intensity and velocity maps for filament A. The panels show,
from left to right: intensity, bisector velocities for 90\% relative
intensity for Fe{\small I} 7090 \AA: original and degraded resolution,
and bisector velocities at 90$\%$ relative intensity for C{\small I}
5380 \AA: original and degraded resolution.}
\label{fig:figure5}
\end{figure}

\begin{figure}
\centering
\includegraphics[width=.5\textwidth,angle=0]{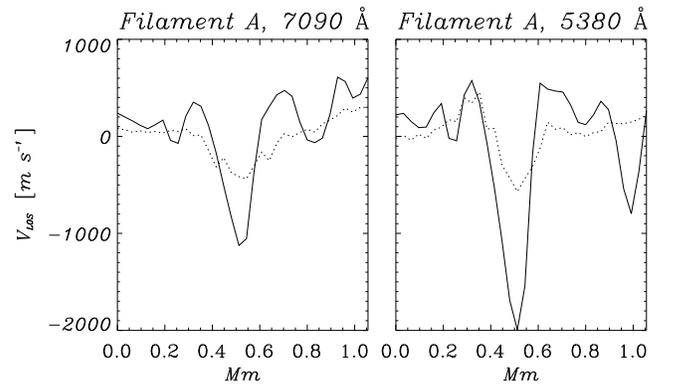}
\caption{Velocity profiles along a horizontal cut perpendicular to
  filament A at 1.5 Mm, indicated on the intensity map in
  Figure~\ref{fig:figure5}. Shown are bisector velocities for 90$\%$
  relative intensity at original resolution (solid lines) at and
  degraded resolution (dotted lines).}
\label{fig:figure6}
\end{figure}

\begin{figure}
\centering
\includegraphics[width=.5\textwidth,angle=0]{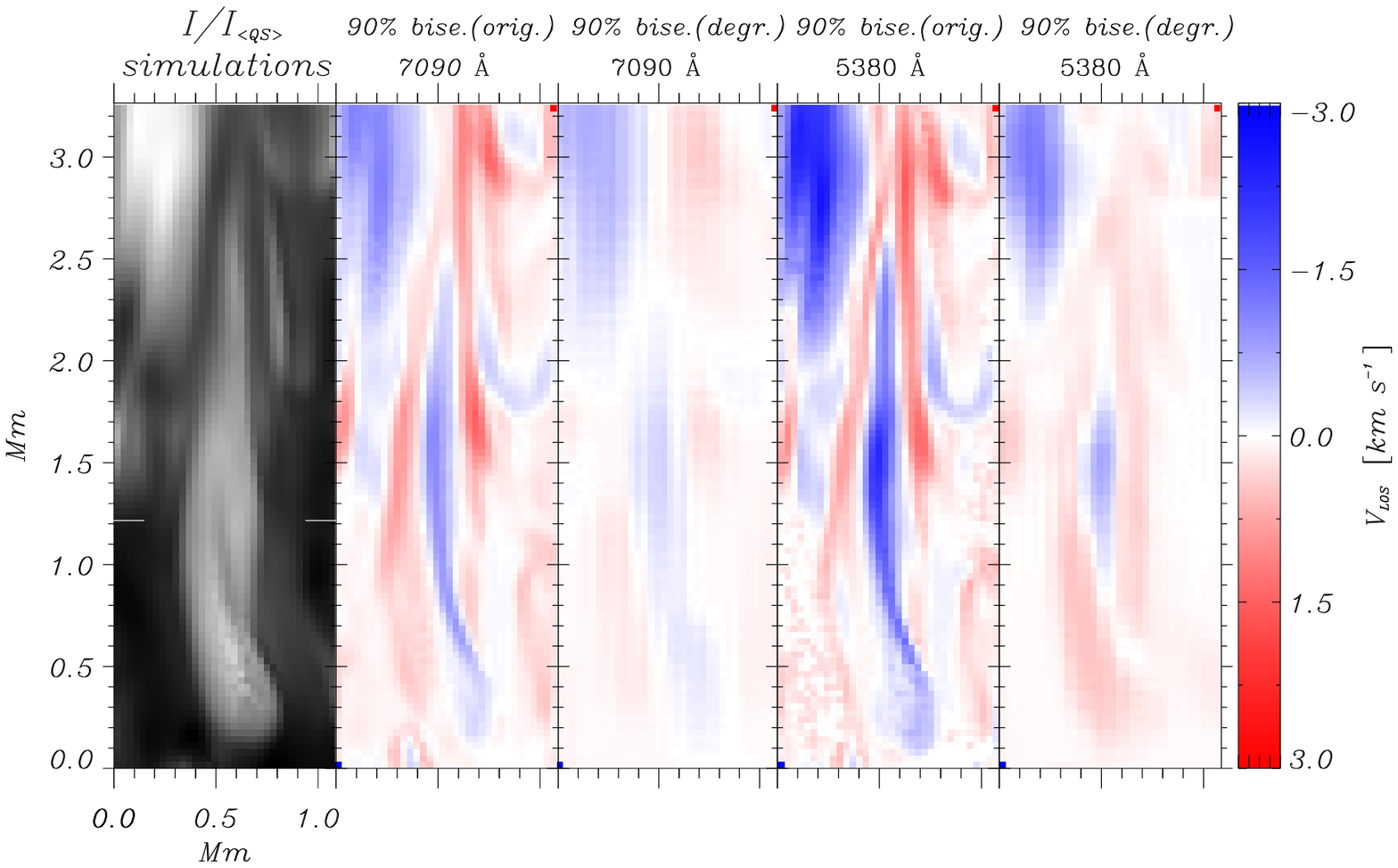}
\caption{Same as Figure~\ref{fig:figure5}, but for filament B.}
\label{fig:figure7}
\end{figure}

\begin{figure}
\centering
\includegraphics[width=.5\textwidth,angle=0]{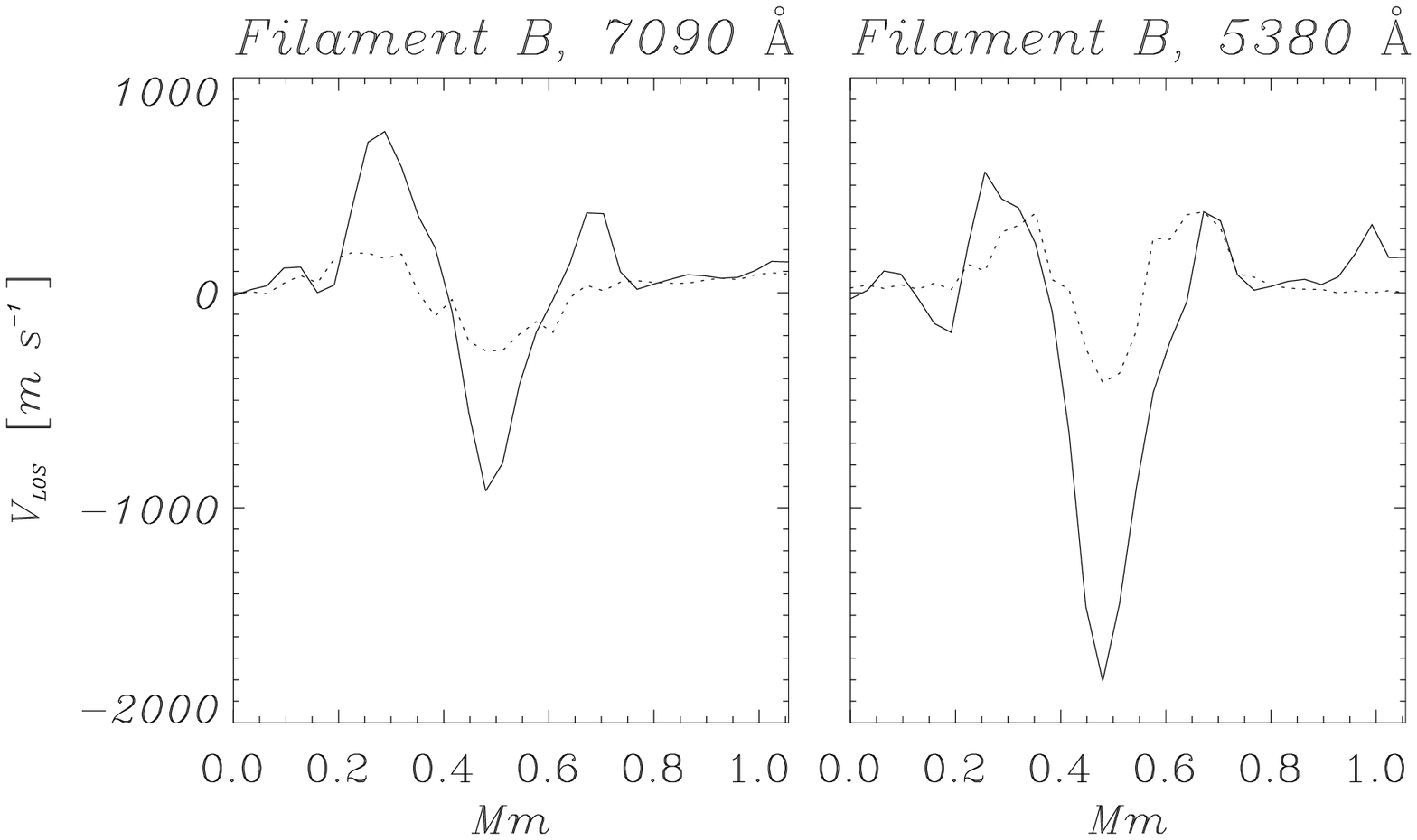}
\caption{Same as Figure~\ref{fig:figure6}, but for filament B.}
\label{fig:figure8}
\end{figure}

\begin{figure}
\centering
\hspace{5mm}
\includegraphics[width=.5\textwidth,angle=0]{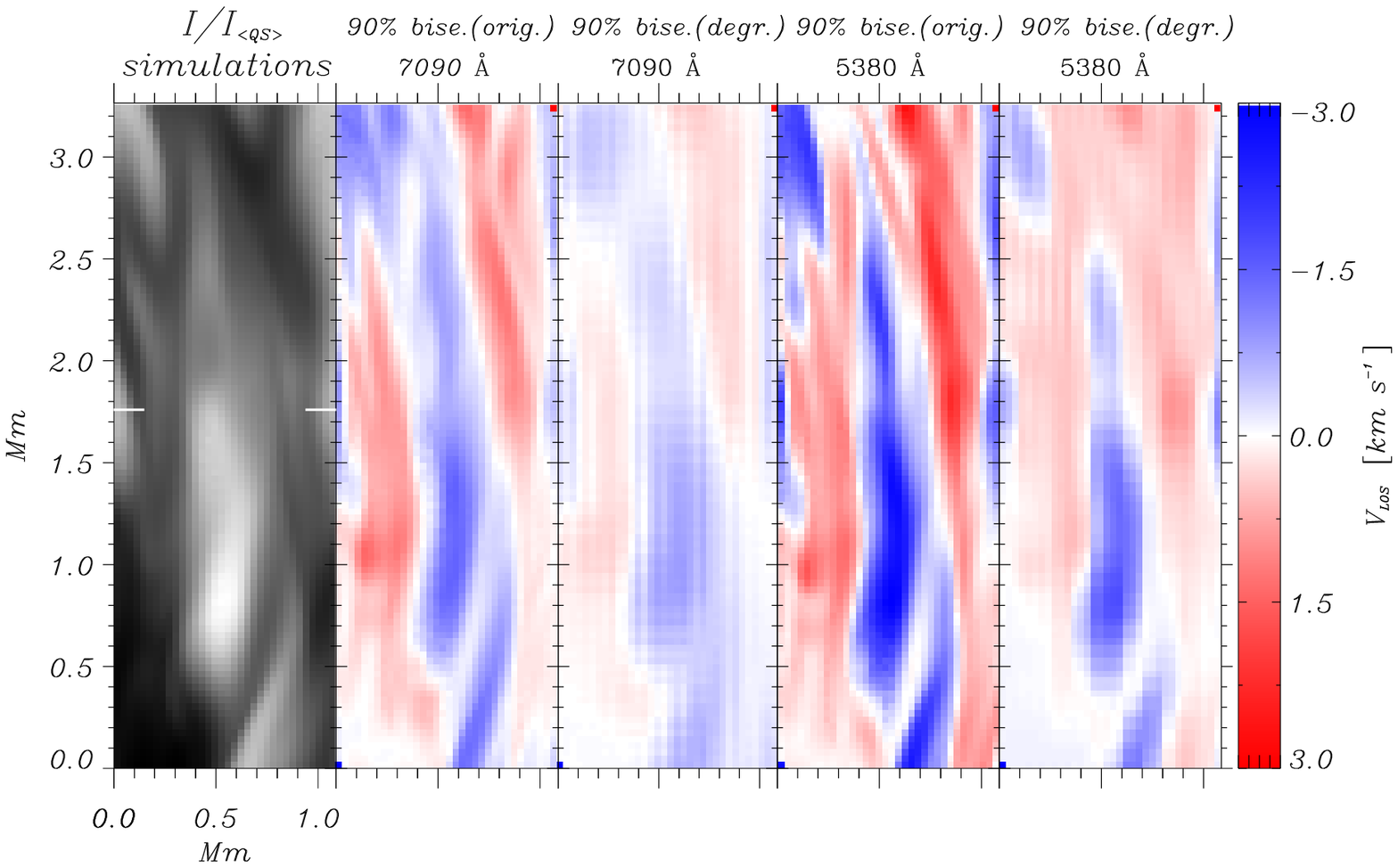}
\caption{Same as Figure~\ref{fig:figure5}, but for filament C.}
\label{fig:figure9}
\end{figure}

\begin{figure}
\centering
\includegraphics[width=.5\textwidth,angle=0]{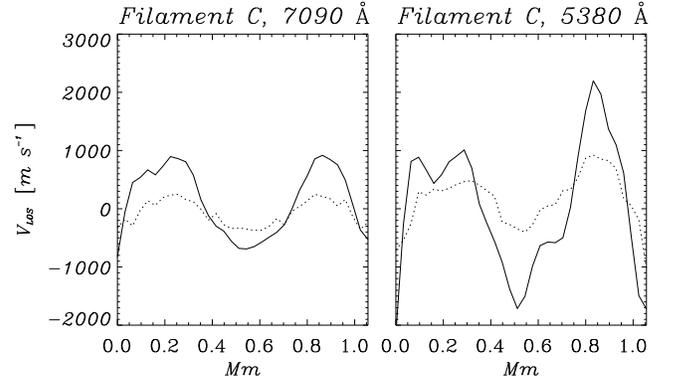}
\caption{Same as Figure~\ref{fig:figure6}, but for filament C.}
\label{fig:figure10}
\end{figure}

\subsection{Line-of-sight velocities of individual filaments}
\label{subsec:filaments}

We have selected three well-developed penumbral filaments in the
simulation snapshot for a more detailed study of their velocity
structure. The filaments A, B, and C are indicated in the upper left
panel of Figure~\ref{fig:figure1}. The actual vertical velocities from
the simulation in vertical cuts perpendicular to the axes of the three
filaments are shown in Figure~\ref{fig:figure4}. The upflows in the
filaments and the downflows at their edges are well visible. The
overplotted iso-$\tau$ lines show that the maximum speeds are found at
or below optical depth unity, so that they are best sampled by spectral
lines with a low formation height.

The following figures show how the vertical flows in these filaments
and their immediate surroundings are represented by the velocities
determined from the line bisectors. Figure~\ref{fig:figure5} shows
maps of the continuum intensity and the 90\% bisector velocities for
the two lines and at both original and degraded resolution for
filament A.  The velocity maps demonstrate the strong effect of the
degradation on the detectability of the vertical flows: while the
typical velocity pattern with upflow in the filament and downflow
along its edge is well represented at original resolution, it is
difficult to see in the degraded velocity map from the 7090 \AA~
line. For the more deeply originating 5380 \AA~ line, a somewhat
stronger velocity signal remains even for degraded resolution.  This
is made more quantitative in Figure~\ref{fig:figure6}, which gives the
bisector velocities along a horizontal `slit' indicated by the two
horizontal line segments in the continuum map in
Figure~\ref{fig:figure5}. While the central upflow remains detectable
even for the degraded map from the 7090 \AA~ bisector, the weaker
downflows are reduced to values around or below the detection limit.
  Similar results are found for other cuts perpendicular to the
  filament.  For the 5830 \AA~ line, the velocity signal is stronger
and the downflow remains detectable for degraded resolution, at least
at one side of the filament.

Qualitatively similar results are obtained for the other two filaments
(Figures~\ref{fig:figure7}--\ref{fig:figure10}): the velocity signal
is so strongly weakened by degradation for the 7090 \AA~ line that the
downflow signal at the filament edges is reduced to an amplitude below
the typical observational detectability limit of
$\sim$200~m$\,$s$^{-1}$ in most cases.  For the 5380 \AA~ line,
the situation is somewhat better: the deep origin of the line and the
lower degree of spatial smearing (owing to the shorter wavelength)
leave detectable velocity signals for both upflows and downflows, even
after degradation.

\subsection{Effect of viewing angle}
\label{subsec:incl}

When a sunspot is observed away from disk center, the observational
detection of vertical velocities in penumbrae is complicated by the
projection of the strong Evershed flow onto the direction of the
LOS. This is relevant also for the results of Bellot Rubio et
al. (2010), who studied a sunspot located only $5.4^\circ$ away from
disk center. Since $\sin 5.4^\circ\simeq0.094$, the component of a
horizontal flow along the inclined LOS can reach a value of up to
$\sim10\%$ of the horizontal flow speed, depending on the angle
between the velocity vector and the direction toward disk center. With
a horizontal Evershed flow of several km$\,$s$^{-1}$, this leads to
projected LOS velocities of several hundred m$\,$s$^{-1}$ for
penumbral filaments directed towards (or away from) disk center. For
the observations presented by Bellot Rubio et al. (2010) the
projection of the Evershed flow leads to a blueshift, i.e., apparent
upflows if (wrongly) interpreted as vertical flows. Their Figure 3
shows a clear trend toward increasing overall blueshift as the
penumbral filaments become more aligned with the slit, which in turn
is nearly parallel to the direction toward disk center. In fact, their
bisectors at the 80-88\% relative intensity level are all blueshifted,
consistent with the maximum of the Evershed flow in the deeper layers.

\begin{figure*}
\centering
\includegraphics[width=1.0\textwidth,angle=0]{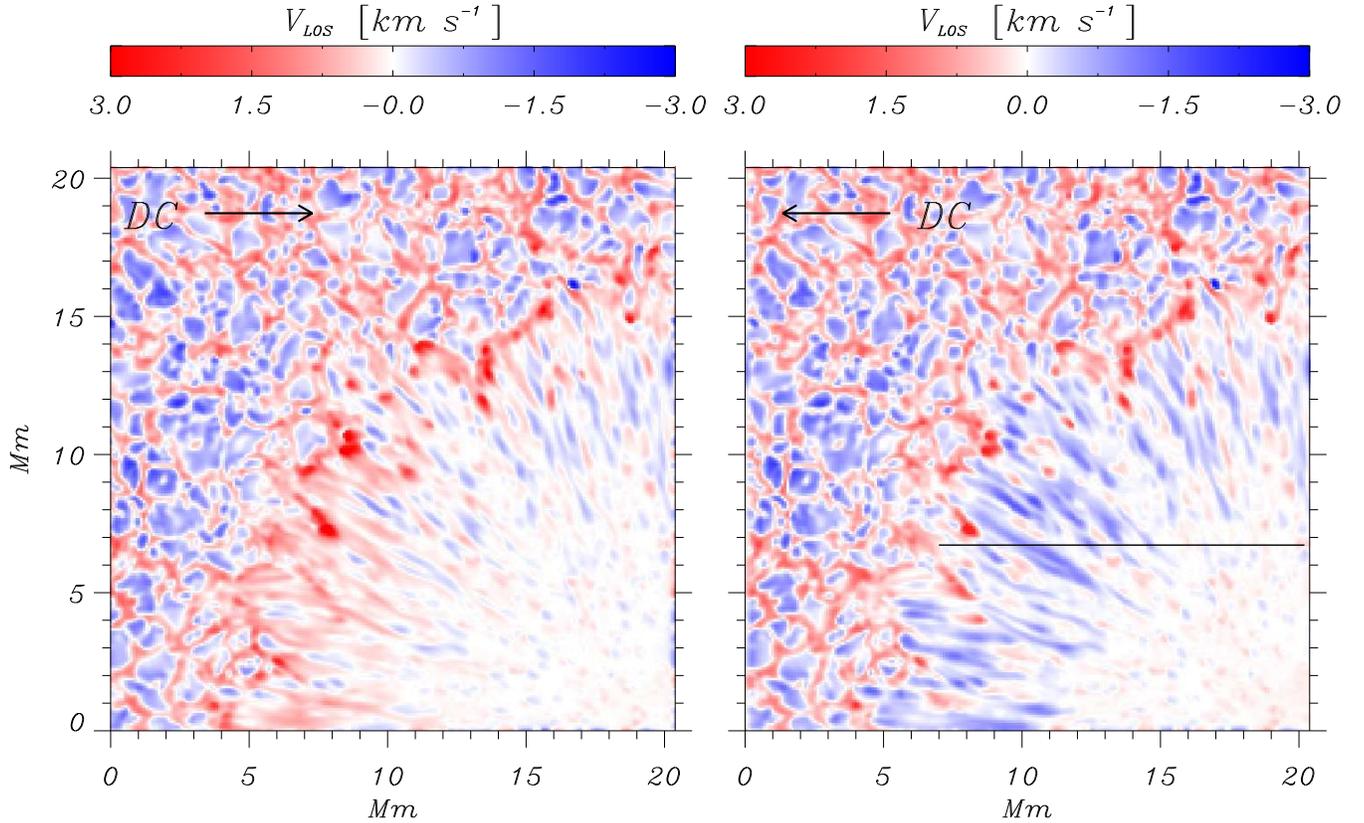}
\caption{LOS velocity maps derived from the bisector shift at 90\%
  relative intensity of Fe{\small I} 7090 \AA~ for viewing angles inclined
  to the vertical by 5.4$^\circ$.  Left panel: disk center (DC) to the
  right; right panel: disk center to the left.}
\label{fig:figure11}
\end{figure*}

\begin{figure}
\hspace{-3mm}
\includegraphics[width=0.5\textwidth,angle=0]{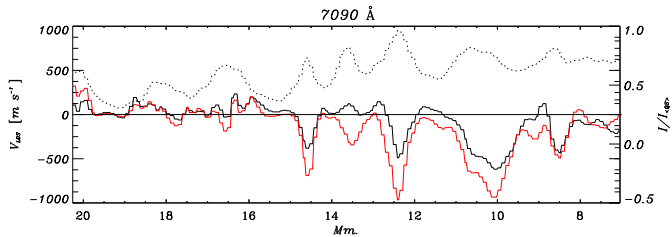}
\caption{Profiles of the LOS velocity for a viewing angle of 5.4$^\circ$
  along the `slit' indicated on the right panel of Figure~\ref{fig:figure11}.
  Note that we reversed the horizontal direction along the `slit'
  compared to Figure~\ref{fig:figure11} in order to present quantities
  directly comparable to Figure 3 in Bellot Rubio et al. (2010).
  LOS velocities were derived from bisector
  shifts of Fe{\small I} 7090 \AA. The black solid line corresponds to
  the average bisectors for 0-8\% relative intensity (line core) and the
  red line to average bisectors for 80-88$\%$ relative intensity (line
  wing).  The dotted line shows the profile of the continuum intensity
  at 7090 \AA.}
\label{fig:figure12}
\end{figure}

To study the effect of the finite viewing angle on the bisector
velocities from the simulated sunspot and to directly compare with the
results of Bellot Rubio et al. (2010), we have redone the line synthesis
assuming an inclination by $5.4^\circ$ to the vertical.
Figure~\ref{fig:figure11} shows the resulting spatially and spectrally
degraded maps of the LOS velocity determined from the bisector shifts of
the 7090 \AA~ line at 90\% relative intensity level. Disk center is
located to the left or to the right, respectively, in the direction of
the slit indicated in the right panel.  In the latter case (right
panel), which corresponds to the observations of Bellot Rubio et
al. (2010), the penumbral region is clearly dominated by blueshifts due
to the projection of the Evershed flow. LOS velocity profiles along the
slit are shown in Figure~\ref{fig:figure12}.  As the penumbral filaments
become more aligned with the slit, the projection of the Evershed flow
onto the direction of the LOS increases. In addition, the Evershed flow
speed increases outward in the penumbra. The combination of these two
effects leads to the downward tilt in the velocity profile shown in
Figure~\ref{fig:figure12}, which is
more conspicuous for the bisectors in the line wings since the Evershed
flow is faster in the deeper layers. These properties of the
velocity profiles are very similar to those found by Bellot Rubio et
al. (2010, see their Figure 3), strongly suggesting that both result
from the projection of the Evershed flow. The difference between the
bisector shifts of line core and line wings is even
higher in the observations, indicating that the increase of the Evershed
flow with depth is possibly underestimated in the simulations.

\section{Discussion and conclusion}

We have found that the observational detection of overturning
convection in penumbral filaments can be severely compromised by
a) finite spatial and spectral resolution and b) projection of the
Evershed flow for observations not taken exactly at disk center. In
particular, the relatively weak downflows along the edges of filaments
suffer most from these effects: for the 7090~\AA~ line their
observational signal is reduced to values below the limit of reliable
detection around $\sim200$~m$\,$s$^{-1}$ in most cases. The
situation is somewhat better for the 5380 \AA~ line, which samples the
higher velocity amplitudes in the deeper layers of the atmosphere
owing to its lower formation height. Obviously, better spatial
resolution also significantly improves the detectability of these
small-scale velocity structures.

The combined effects of degradation and projection probably also
  affected the observations of Bellot Rubio et al. (2010), so that
they cannot decide upon the presence or absence of overturning
convection in the penumbra exceeding the stated detection limit of
 $\sim50$~m$\,$s$^{-1}$. Note that our analysis did not take into
account the effects of noise (due to the detector, electronics, and
seeing) in the observed line profiles.  Comparing our degraded images
with Figure 2 of Bellot Rubio et al. (2010) also indicates that the
spatial degradation of the observations is stronger than in our
attempts to simulate them.  However, the main problem of these
observations is the projection of the Evershed flow onto the LOS. Even
at a heliocentric angle of only 5.4$^\circ$, up to 10\% of the
Evershed flow is projected onto the LOS and the trend in the velocity
profile shown in Figure 3 of Bellot Rubio et al. (2010) clearly
indicates that this effect is present in their observations. This is
also supported by the fact that the bisector blueshifts are
significantly stronger for the line wings than for the line core.
Therefore, the absence of redshifted bisectors must not
  necessarily be taken as evidence for the absence of downflows since
the projection of the Evershed effect easily leads to a blueshift of
several 100~m$\,$s$^{-1}$.

  On the other hand, the non-detection of overturning penumbral
  convection by Franz \& Schlichenmaier (2009) and Bellot Rubio et
  al. (2010) could, in principle, indicate that these flows are
  overestimated in current numerical simulations owing to insufficient
  spatial resolution, artificial boundary conditions, and possibly
  insufficient domain depth. We have recently carried out a convergence
  study (Rempel, in preparation) for which the numerical grid spacing
  was varied between 96~km and 16~km in the horizontal directions and
  between 32~km and 12~km in the vertical direction.  The simulation
  analyzed here is the second best resolved simulation of that series
  (the highest resolution case has not yet been computed with non-grey
  radiative transfer).  Since we do not use explicit diffusivities,
  changes in the grid spacing directly affect the overall numerical
  dissipation (which scales at least linearly with grid spacing near
  discontinuities but with a higher order in well-resolved regions).
  The amount of overturning motions (characterized by the vertical rms
  velocity at $\tau=1$) turns out to be robust: it is directly tied to
  the penumbral brightness, which does not change significantly with
  resolution. On the other hand, we find that the average width of
  filaments decreases somewhat with increasing resolution since they are
  still only marginally resolved. It is therefore well conceivable that
  our current simulation overestimates the visibility of convective
  motions in the penumbra. We also investigated the influence of the top
  boundary condition. While the overall extent of the penumbra depends
  on the choice of the boundary condition, the detailed structure of the
  magnetoconvection is mostly unaffected. We find an approximate
  relationship of the form $I\propto\sqrt{v_{z}^{\mbox{rms}}(\tau=1)}$
  between the azimuthally averaged bolometric intensity and the rms
  vertical velocity defined through the azimuthal average at each radial
  position of the spot (Rempel 2011), independent of the boundary
  condition and extent of the penumbra.  As a consequence, the
  prediction that the vertical rms velocity in the penumbra should be
  about half of the value found in quiet sun is fairly robust; it is
  also consistent with the width of distribution functions inferred from
  Hinode data by Franz \& Schlichenmaier (2009).

Our analysis indicate that the 5380 \AA~ line is better suited than the
7090 \AA~ line for the detection of overturning convection in penumbral
filaments. This is due to the fact that this line originates deeper in
the atmosphere (and thus samples higher velocities) and that the shorter
wavelength affords a higher spatial resolution for a given
telescope. Even after degradation, many downflows exceed a typical
detectability threshold of 200~m$\,$s$^{-1}$. Results on penumbral
up- and downflows using the 5380 \AA~ line were obtained by
Schlichenmaier \& Schmidt (1999) with the German VTT on Tenerife and
recently by Joshi et al. (2011) and Scharmer et al. (2011) with the CRISP 2D spectropolarimeter at
the SST.  In any case, such observations are complicated by the
projection of the strong Evershed flows and therefore should be carried
as near as possible to disk center -- assuming that the Evershed flow is
exactly horizontal, which obviously will not always be the case.

\acknowledgments

Discussions with Drs. J. Hirzberger and A. Lagg are gratefully
acknowledged. The authors thank Dr. B. Lites for helpful comments on
the manuscript. LB is grateful to Inter-University Centre for
Astronomy and Astrophysics (IUCAA) Reference Center at Department of
Physics, Mohanlal Sukhadia University, Udaipur, India, for providing
computational facilities during his visit.

The National Center for Atmospheric Research (NCAR) is sponsored by the
National Science Foundation. Computing resources for the sunspot model
utilized in this investigation were provided through NCAR's Computational
Information Systems Laboratory (CISL) as well as through the NSF Teragrid
at the National Institute for Computational Sciences (NICS) under grant TG-AST100005.


\end{document}